\newcommand{\CLASSINPUTtoptextmargin}{2.4cm}
\newcommand{\CLASSINPUTbottomtextmargin}{4.6cm}
\documentclass[conference]{IEEEtran}
\IEEEoverridecommandlockouts
\usepackage{authblk}
\usepackage{cite}
\usepackage{amsmath,amssymb,amsfonts}
\usepackage{amsmath}  
\usepackage{algorithm}  
\usepackage{algorithmicx}  
\usepackage{algpseudocode}  
\usepackage{amsfonts}  
\usepackage{graphicx}
\usepackage{textcomp}
\usepackage[table]{xcolor}
\usepackage{tabularx}
\usepackage{url}
\usepackage[bookmarks=false]{hyperref}
\usepackage{rotating}
\usepackage{subcaption}
\usepackage{multirow}
\usepackage{pifont}
\usepackage{tcolorbox}
\definecolor{tumColorLightBlue}{HTML}{f0f5fa} 
\usepackage{cleveref}

\usepackage{tablefootnote}
\usepackage{threeparttable}
\usepackage{multirow}
\usepackage{booktabs}
\usepackage{harveyballs}
\renewcommand{\baselinestretch}{0.96}
\usepackage{eso-pic}
\usepackage[outdir=./]{epstopdf}
\usepackage{fancyhdr}

\def\BibTeX{{\rm B\kern-.05em{\sc i\kern-.025em b}\kern-.08em
    T\kern-.1667em\lower.7ex\hbox{E}\kern-.125emX}}

\AddToShipoutPictureBG*{%
  \AtPageUpperLeft{%
    \put(0.07\paperwidth,-1.2cm){%
      \fbox{%
        \begin{minipage}{\textwidth}
          \centering
          \small
          \textbf{This paper has been accepted for publication in the proceedings of the IEEE International Conference on Communications (ICC), June 2025. The copyright for this paper will be transferred to IEEE. ©2023 IEEE. Personal use of this material is permitted. Permission from IEEE must be obtained for all other uses, in any current or future media, including reprinting/republishing this material for advertising or promotional purposes, creating new collective works, for resale or redistribution to servers or lists, or reuse of any copyrighted component of this work in other works.}
        \end{minipage}
      }%
    }%
  }%
}

\begin{document}
\title{Learning-Based Traffic Classification for Mixed-Critical Flows in Time-Sensitive Networking\vspace{-0.2cm}}

\author[1]{Rubi Debnath}
\author[2,*]{Luxi Zhao\thanks{* Luxi Zhao is the corresponding author.}}
\author[3]{Sebastian Steinhorst\vspace{-0.2cm}}

\affil[1,3]{TUM School of Computation, Information and Technology, Technical University of Munich, Germany}
\affil[2]{Electronic Information Engineering College, Beihang University, Beijing, China}
{
    \makeatletter
    \renewcommand\AB@affilsepx{, \protect\Affilfont}
    \makeatother
    \affil[1,3]{firstname.lastname@tum.de}
    \affil[2]{zhaoluxi@buaa.edu.cn}
}
\makeatletter
\patchcmd{\@maketitle}
  {\addvspace{0.5\baselineskip}\egroup}
  {\addvspace{-1\baselineskip}\egroup}
  {}
  {}
\makeatother

\maketitle

\begin{abstract}
Time-Sensitive Networking (TSN) supports multiple traffic types with diverse timing requirements, such as hard real-time (HRT), soft real-time (SRT), and Best Effort (BE) within a single network. To provide varying Quality of Service (QoS) for these traffic types, TSN incorporates different scheduling and shaping mechanisms. However, assigning traffic types to the proper scheduler or shaper, known as Traffic-Type Assignment (TTA), is a known NP-hard problem. Relying solely on domain expertise to make these design decisions can be inefficient, especially in complex network scenarios. In this paper, we present a proof-of-concept highlighting the advantages of a learning-based approach to the TTA problem. We formulate an optimization model for TTA in TSN and develop a Proximal Policy Optimization (PPO) based Deep Reinforcement Learning (DRL) model, called ``TTASelector'', to assign traffic types to TSN flows efficiently. Using synthetic and realistic test cases, our evaluation shows that TTASelector assigns a higher number of traffic types to HRT and SRT flows compared to the state-of-the-art Tabu Search-based metaheuristic method.
\end{abstract}

\section{Introduction}
\label{sec:introduction}
Time-Sensitive Networking (TSN)~\cite{8021Q} is a set of IEEE sub-standards designed to provide deterministic, low and bounded latency over Ethernet networks. TSN aims to meet the stringent requirements of real-time applications across various domains, such as industrial automation, automotive systems, spacecraft, and integrated wired-wireless~\cite{rubi_vtc} communications. With the growing need for reliable and predictable networking in sectors like Industry 4.0 and connected vehicles, TSN represents a significant advancements over traditional best-effort Ethernet. TSN offers various shaping and scheduling mechanisms~\cite{rubi_rtcsa} to ensure that flows are transmitted predictably and reliably, minimizing both jitter and latency. In a mixed-critical TSN network, there are generally three types of traffic: 
\begin{enumerate}
    \item \textbf{Hard-Real Time (HRT):} HRT flows have strict deadlines, where missing a deadline can result in safety-critical issues.
    \item \textbf{Soft-Real Time (SRT):}  SRT flows have soft deadlines, where missing the deadline can degrade performance and cause user dissatisfaction but does not result in safety-critical concerns.
    \item \textbf{Non-Time Critical (NTC):} NTC flows have no timing guarantee requirement.
\end{enumerate}

\begin{figure}[t!]
    \centering
        \includegraphics[scale=0.36,trim={0.5cm 0.1cm 0.5cm 0.1cm}, clip]{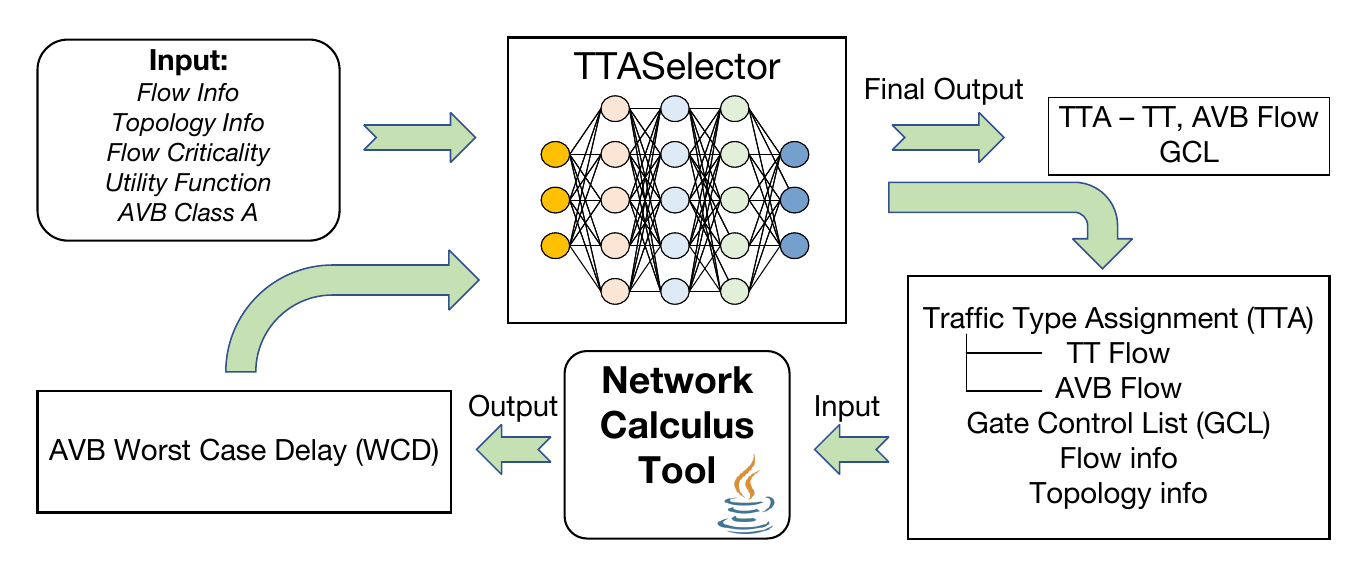}
        \caption{Contribution and overall workflow of the paper.}
    \label{fig:overview}
    \vspace{-0.5cm}
\end{figure}
This classification enables TSN to support a range of applications while ensuring that critical traffic is prioritized appropriately. There are different traffic type in a TSN network, including: Time-Triggered (TT), where the flows are transmitted based on a pre-determined static table known as the Gate Control List (GCL), Audio Video Bridging (AVB), for dynamically scheduled flows, and Best Effort (BE), which does not have any Quality of Service (QoS) requirements. Assigning the appropriate traffic type (TT or AVB), known as Traffic Type Assignment (TTA), to HRT and SRT flows to meet necessary QoS requirements is an NP-hard problem~\cite{voica_traffic_assignment} that significantly impacts network performance. Traditional mathematical solvers struggle to manage large-scale networks and flows, with \cite{ilp_limitations} highlighting the limitations of Integer Linear Programming (ILP) in TSN. To the best of our knowledge, only one prior work addresses TTA in TSN. Gavrilut et al.~\cite{voica_traffic_assignment} proposed a Tabu search-based metaheuristic algorithm to solve the TTA optimization problem. Although metaheuristic algorithms can deliver solutions within reasonable time complexity, this complexity escalates with large-scale networks and increasing number of flows. Deep Reinforcement Learning (DRL) has shown recent success in tackling NP-hard problems in TSN~\cite{ttdeep}. Related works have applied different DRL algorithms to address TSN scheduling and shaping mechanisms, but the TTA problem remains largely unaddressed in the research community. This gap draws attention to the potential of DRL in offering scalable solutions to the TTA challenge in TSN.

\noindent \textbf{Motivation:} Motivated by this challenge, our work presents a framework (refer to Fig.~\ref{fig:overview}) to address the TTA problem in TSN. Our framework takes the network topology, flow information, flow criticality, utility function, and AVB Class (detailed in Section~\ref{sub:utility_func}) as input. Our objective is to ensure that all HRT flows are scheduled while maximizing QoS for SRT flows (detailed in Section~\ref{sec:model}). To achieve this, we propose a DRL-based framework to solve the NP-hard TTA problem in a mixed-critical TSN network containing both HRT and SRT flows. 

\noindent \textbf{Problem Formulation:} Given a network topology $\mathcal{G}$, a set of TSN flows $\mathcal{F}$, flow criticality (HRT, or SRT), the routes of the flow, available AVB class, utility function of the SRT flows, and the flow parameters, our goal is to determine the traffic type of TSN flows such that (1) all HRT flows are schedulable (i.e., they meet their deadlines) and (2) the total QoS of SRT flows is maximized. Our model assigns two types of traffic: TT or AVB. TT flows are transmitted based on the GCL. Therefore, for TT flows, our model computes the GCL based on a simple As-Soon-As-Possible (ASAP) scheduling strategy. The GCL defines the \emph{opening} and \emph{closing} time of the gates of TT traffic. For the AVB traffic, we assign the AVB class and the Credit-Based Shaper (CBS) parameters ($idleSlope$) for the AVB class.

\subsection{Contribution}
\noindent In summary, our main contributions are:
\begin{enumerate}
    \item We introduce a framework (illustrated in Fig.\ref{fig:overview}) for end-to-end TTA with integrated Network Calculus (NetCal) to determine the Worst Case Delay (WCD) for AVB flows (Section~\ref{sec:implementation}).
    \item We formulate an optimization problem based on our framework to maximize the schedulability of HRT flows and improve the QoS of SRT flows, where QoS is modeled using a utility function (refer Section~\ref{sub:utility_func}).
    \item Recognizing the NP-hard nature of this problem, we transform it into a Markov Decision Process (MDP) and propose a Proximal Policy Optimization (PPO) based TTA algorithm, ``TTASelector'', to derive the optimal policy for TTA in TSN networks.
\end{enumerate}

\section{System Model and Problem Formulation}
\label{sec:model} 
In this section, we present the architecture and application models.

\subsection{Architecture Model}
\label{subsec:arch_model}
The TSN network consists of switches $\mathcal{(SW)}$ and end stations $\mathcal{(ES)}$. We represent the TSN network as an undirected graph, denoted as $\mathcal{G(V, E)}$, where $\mathcal{V}$ denotes the nodes or vertices and $\mathcal{E}$ denotes the edges or links connecting the nodes. The node set $\mathcal{V}$ includes both $\mathcal{ES}$ and $\mathcal{SW}$, represented as $\mathcal{V = (ES \cup SW)}$, while $\mathcal{E}$ contains all links between the nodes in the network, where $e_{j}$ $\in$ $\mathcal{E}$, and $j$ indicates the link/edge number. Without loss of generality, we assume that all physical links operate at the same transmission rate, $\mathcal{C}$. Each TSN flow has a single source and can have one or more receivers. The route of the flows are given and fixed in our model. Every TSN switch has eight queues in the egress port, supporting both TT and AVB flows. For simplicity, we consider one queue for TT and one for AVB in the egress port. We consider one AVB Class, namely AVB Class A, with an $idleSlope$ set to 75\%. 

\begin{figure}[t!]
    \centering
    \includegraphics[scale=0.5,trim={0.1cm 0.4cm 0.1cm 0.6cm}, clip]{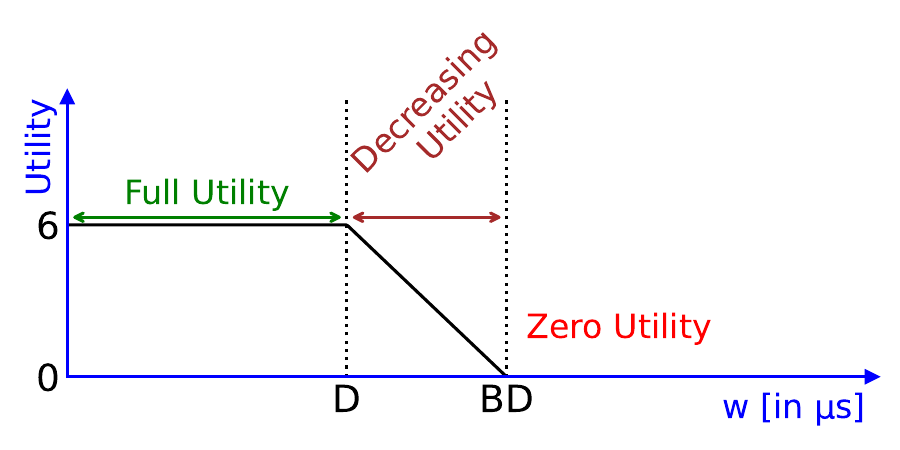}
        \caption{Utility Function (D represents the soft deadline of the SRT flows, and BD represents the buffer deadline).}
    \label{fig:utility_function}
    \vspace{-0.4cm}
\end{figure}

\subsection{Application Model}
\label{sub:app_model}
In the TSN network, we consider two types of flows with distinct timing requirements: HRT flows, which have strict deadlines, and SRT flows, which have soft deadlines. We denote the set of all flows in a network as $\mathcal{F = F^{HRT} \cup F^{SRT}}$, where the two sets $\mathcal{F^{HRT}}$ and $\mathcal{F^{SRT}}$ correspond to the set of all HRT, and SRT flows in the network respectively. Each flow $f_{i}$ consists of the following parameters (refer to Eq.~\ref{eq:flow}). $\forall f_i \in \mathcal{F}$, where $i = 1,2,...,|\mathcal{F}|$ 
\begin{equation}
    f_i = \langle f_i.src, f_i.dst, f_i.prd, f_i.pld, f_i.dln, f_i.crc \rangle,
    \label{eq:flow}
\end{equation}
\noindent where, $f_i.src$ is the source, $f_i.dst$ is the destination, $f_i.prd$ is the periodicity in $\mu$s, $f_i.pld$ is the payload in Bytes, $f_i.dln$ is the hard or soft deadline in $\mu$s, and $f_i.crc$ is the criticality of the flow (HRT or SRT). HRT flows have strict deadlines, and missing them may lead to safety risks. By contrast, SRT flows have soft deadlines, and missing these deadlines results in performance degradation. Each SRT flow has its own utility function~\cite{voica_traffic_assignment} which is given as an input (shown in Fig.~\ref{fig:utility_function}). For SRT flows, we define an additional buffer period following the soft deadline, referred to as the buffer deadline (BD). If the WCD of an SRT flow exceeds the soft deadline but falls within the BD, the flow is considered scheduled, although, the performance is degraded. The WCD is defined as the maximum end-to-end delay of the flows in the TSN network. WCD is a crucial performance indicator, as it ensures timing guarantees in a TSN network. For the TT flows, the WCD is determined by the GCL, and for the AVB flows the WCD is determined by the NetCal tool. To model the QoS and to quantify the performance degradation of SRT flows, we use the utility function~\cite{voica_traffic_assignment}, as depicted in Fig.~\ref{fig:utility_function}. 

\subsection{Utility Function}
\label{sub:utility_func}
We define the utility~\cite{voica_traffic_assignment} function of SRT flows, illustrated in Fig.~\ref{fig:utility_function}, where $D$ denotes the soft deadline, and BD denotes the buffer deadline for SRT flows. The utility function, denoted as $U_i(w)$, quantifies the QoS for each SRT flow. The utility value is initially set to a maximum of 6 and remains constant up to the soft deadline, $D$. Once $D$ is exceeded, the utility begins to decrease until it reaches zero at BD. In our paper, we set the value of BD to 1.5$\times$D. The initial utility value of 6 can be adjusted by system engineers to optimize the model for different applications. Similarly, BD is derived from $D$ and remains configurable based on system requirements. The utility function is defined as:
\begin{equation}
U_i(w) = 
\begin{cases} 
s, & w \leq D \\
-\frac{12}{D} \cdot (w - D) + 6, & D < w \leq BD \\
0, & w > BD
\end{cases}
\label{eq:utility}
\end{equation}
\noindent where D is the deadline, $s$ is the utility value, and $w$ represents the WCD of the AVB flow. In our paper, we set the value of $s$ to 6, and BD to $1.5 \times D$. Finding the optimal utility or BD value and discussing the impact of the different utility or BD value is out of the scope of this paper. For proof of concept, we have chosen these values.

\subsection{Input and Output Files}
\noindent \textbf{Input:} The model requires the following inputs: (1) the network topology, (2) the set of flows $f_i \in \mathcal{F}$, (3)  the type of flow (HRT or SRT), (4) the utility function of SRT flows, and (5) the available AVB Class. 

\noindent \textbf{Output:} The model outputs the traffic type assignment for the given network, denoted as $\mathcal{TTA}$. Additionally, it generates the GCL for TT flows. For AVB flows, the model integrates with the NetCal tool to determine the WCD of AVB flow. The complete framework is shown in Fig.~\ref{fig:overview}. It is important to note that, running the model requires an initial solution, which is based on a naive approach and is discussed in Section~\ref{sub:initial_sol}.

\subsection{Gate Control List (GCL)}
Several studies explore algorithms for generating GCL schedules~\cite{gate_array, window_niklas, ttdeep}, with surveys available on various GCL approaches in~\cite{survey_gcl_access, survey_gcl_rtas}. In this paper, we use the ASAP algorithm to generate the GCL of the TT flows. The algorithm further calculates the offset for each TT flow, where the offset represents the time at which a TT flow is sent from the source node. 

\noindent\textbf{ASAP Approach:} To generate the GCL using the ASAP approach, we divide the hyperperiod (also known as the scheduling cycle), denoted by $H$, into discrete time slots. Following similar approaches in~\cite{nayak, ttdeep}, we set the granularity of each time slot to $\frac{1}{64}$ms = 15.625$\mu$s. This time slot allows the transmission of 1500 Bytes (one Maximum Transmission Unit (MTU)) at a link speed of 1 Gbps. We calculate the $H$ as follows:
\vspace{-0.2cm}
\begin{equation}
    H = \mathrm{LCM}(f_i.prd), \;\; \forall f_i \in \mathcal{F}.
\end{equation}
The algorithm then assigns offsets and selects time slots across each link in the route for each TT flow. The ASAP algorithm takes the network topology, TTA assignment, TT flows, and routes of the TT flows as the input. The output of the ASAP algorithm is the GCL of the TT flows. For the first iteration, the ASAP algorithm takes the initial traffic assignment based on the naive approach denoted as $\mathcal{TTA}^0$ (given in Algorithm~\ref{alg:initial_solution}).

\subsection{Constraints}
\label{sub:constraints}
We outline the primary constraints required for GCL generation:

\noindent \textbf{Link Constraint:} No two TT flows should be scheduled on the same link within the same time slot. Additionally, the combined number of TSN flows (both TT and AVB) transmitted over a link must remain within the link capacity ($C$) to prevent queue backlog.

\noindent \textbf{Offset Constraint:} The offset of a TT flow must be less than its periodicity.

\noindent \textbf{Deadline Constraint for HRT flows:} For every flow $f_i \in F^\mathcal{HRT}$, the worst case delay of the flow ($WCD(f_i)$) must not exceed its deadline.

\subsection{Worst Case Delay (WCD) Analysis}
\label{sec:wcd}
The WCD for TT traffic is defined by the GCL and the selected offset for each flow. However, the WCD for AVB traffic is not known immediately. To determine this, we employ the NetCal tool from~\cite{luxi_netcal_rtas}, which evaluates TT and AVB traffic within the same network. The WCD obtained for AVB flows is subsequently used in our model to compute the utility value of each respective AVB flow (refer Fig.~\ref{fig:overview}). 

\subsection{Optimization Function}
\label{sub:optimization}
The objective of the optimization function is to assign the traffic type to each TSN flow such that: (1) all HRT flows are schedulable, and (2) the total utility of SRT flows is maximized. We define $SCH_{f_i}$ as a function indicating whether flow $f_i$ is schedulable. The schedulability of HRT flows is maximized by:
\vspace{-0.2cm}
\begin{equation}
\delta_{HRT} = \max\sum_{f_i \in \mathcal{F}^{HRT}} SCH_{f_i}
\end{equation}
The total utility ($\delta_{SRT}$) of the SRT flows is maximized as follows:
\begin{equation}
\delta_{SRT} = \max \sum_{f_i \in \mathcal{F}^{SRT}} U_{i} (WCD(f_i))
\label{eq:srt_utility}
\end{equation}
Eq.~\ref{eq:srt_utility} maximizes the total utility $\delta_{SRT}$ of the SRT flows by mapping the WCD of the flow to a utility value using the utility function $U_i$ (refer to Eq.~\ref{eq:utility}). 

\begin{algorithm}[!t]
    \caption{InitialSolution $(\mathcal{G(V,E)}, \mathcal{F})$}
    \begin{algorithmic}[1]
        \State \textbf{Initialize} $\mathcal{TTA}^0 \gets \phi$ \Comment{Initial TTA solution to empty mapping}
        \ForAll{$f_i \in \mathcal{F}$}
            \If{$f_i.\text{crc} = \text{SRT}$}
                \State $\mathcal{TTA}^0(f_i) \gets \text{AVB}$
                \State $f_i.\text{AVB} \gets \text{Class\;A}$
            \ElsIf{$f_i.\text{crc} = \text{HRT}$}
                \State $\mathcal{TTA}^0(f_i) \gets \text{TT}$
            \EndIf
        \EndFor
        \State $GCL \gets \text{ASAP}(\mathcal{G(V,E)}, \mathcal{TTA}^0, \mathcal{F})$ \Comment{Generate the GCL for the TT flows}
        \State $WCD(AVB) \gets \text{NetCal}(GCL, \mathcal{TTA}^0, \mathcal{G(V,E)}, \mathcal{F})$ \Comment{Find the WCD of the AVB flows using the NetCal tool}
        \State \textbf{return} $\mathcal{TTA}^0, GCL$
    \end{algorithmic}
    \label{alg:initial_solution}
\end{algorithm}
\setlength{\textfloatsep}{10pt}
\setlength{\floatsep}{10pt}

\section{DRL Model}
\label{sec:implementation}
We propose a Proximal policy optimization (PPO)-based model, called ``TTASelector'', for solving the traffic type assignment problem in TSN networks. PPO is widely used due to its training stability by limiting large policy updates~\cite{ppo} leading to a faster and efficient training. 

\noindent \textbf{General Description:} The PPO-based TTASelector model consists of several key components: (1) \textbf{Actor}: The actor is responsible for performing actions — in this model, it assigns the traffic type (TT or AVB) to flows in the network. (2) \textbf{Critic}: The critic evaluates the effectiveness of the chosen traffic type in maximizing HRT scheduling and SRT utility. (3) \textbf{State}: The state represents the current network properties, such as the available link capacity, available slots, and scheduled flows. (4) \textbf{Action}: Actions in this model involve assigning a traffic type, either TT or AVB, to each flow. (5) \textbf{Reward}: The reward function is designed to guide the agent toward scheduling all HRT flows while maximizing the utility of SRT flows. 

\subsection{Initial Solution:}
\label{sub:initial_sol}
To address the TTA problem, we first need an initial solution, denoted as $\mathcal{TTA}^0$, to begin our model training. An initial solution is essential to generate the WCD of AVB flows using the NetCal tool. Therefore, we generate the initial solution using a naive approach (refer to Algorithm~\ref{alg:initial_solution}) by assigning AVB Class A traffic to all SRT flows (Lines 3-5) and TT traffic to all HRT flows (Lines 6-8). Line 10 generates the schedule table (GCL) for the TT flows by using the ASAP approach. The initial TTA solution, the topology, and the flow information is given as an input to the ASAP approach (Line 10). The NetCal tool calculates the WCD of AVB flows (Line 11), which is subsequently used in the DRL model to calculate the schedulability and utility values for AVB flows. Based on the total utility value, the reward is calculated in the DRL model. We only consider AVB Class A traffic in our model. Therefore, we assign AVB class A (Line 5) to the flow.

\subsection{State Space}
\label{sub:state}
The state space, $s_t$ contains information on the network topology, flows, and remaining link capacity ($C$) at any given time step $t$. We represent the state space as $s_t$ = ${s_t^{flow} \cup s_t^{eCap}}$, where $s_t^{flow}$ includes flow information (as described in Eq.~\ref{eq:flow}), buffer deadlines for SRT flows, and the $GCL$. The link state, $s_t^{eCap}$, represents information across all links in the network, where $s_t^{eCap}$ = $\{{s_t^{e_1},s_t^{e_2}, ..., s_t^{e_k}}\}$ and each $e_1, e_2, ..., e_k \in e$ denotes a unique link. $s_t^{eCap}$ provides information on slot availability across the $H$, including both available and occupied slots. This link state can be represented as a matrix, as shown in Eq.~\ref{eq:sedge}. 
\begin{equation}
s_t^{eCap} = 
\left\lbrace 
\begin{array}{cccc}
\xi_{T_1}^{e_1}(t), & \xi_{T_1}^{e_2}(t), & \dots, & \xi_{T_1}^{e_k}(t) \\
\xi_{T_2}^{e_1}(t), & \xi_{T_2}^{e_2}(t), & \dots, & \xi_{T_2}^{e_k}(t) \\
\vdots & \vdots & \ddots & \vdots \\
\xi_{T_n}^{e_1}(t), & \xi_{T_n}^{e_2}(t), & \dots, & \xi_{T_n}^{e_k}(t) \\
\end{array} 
\right\rbrace,
\label{eq:sedge}
\end{equation}
\noindent where $\xi_{T_i}^{e_k}(t)$ denotes the remaining link capacity of link/edge $e_k$ at time interval $T_i$ during time step $t$.

\subsection{Action Space}
The action space in our ``TTASelector'' model is discrete and is defined as follows: 
\begin{equation}
    a_{t} = 
    \begin{cases}  0, & \text{if $f_i$ is assigned as TT}, \\
    1, & \text{if $f_i$ is assigned as AVB},
  \end{cases}
\end{equation}
\noindent where each action $a_t$ represents the assignment of a traffic type to flow $f_i$ during time step $t$.

\begin{table}[!t]
\centering
\renewcommand{\arraystretch}{1.6}
\caption{Reward Function.}
\begin{tabular}{|l|l|l|l|l|}
\hline
    & \shortstack{TT-\\Successful}& TT-Fail   & \shortstack{AVB-\\Successful} & AVB-Fail  \\ \hline
HRT & r$_{HRT,succ}$ & \shortstack{r$_{HRT,fail}$ \\$+ STOP$} & \shortstack{r$_{HRT,succ}$ \\+ $\alpha_{bonus}$} & r$_{HRT,fail}$ \\ \hline
SRT & \shortstack{r$_{SRT,succ}$ \\+ $\beta_{bonus}$} & r$_{SRT,fail}$ & r$_{SRT,succ}$         & r$_{SRT,fail}$ \\ \hline
\end{tabular}
\label{tab:reward}
\end{table}

\subsection{Reward}
\label{sub:reward}
We define eight distinct reward scenarios in our model, differentiating rewards based on the flow type: HRT and SRT. For each flow type, the agent has two possible actions: assigning traffic type TT or AVB. Each action outcome can either be a successful scheduling or a failure. Table~\ref{tab:reward} summarizes the reward strategy of our model. The TTASelector aims to assign HRT flows to the AVB traffic type whenever possible, provided that the scheduling is successful. Therefore, a positive bonus ($\alpha_{\text{bonus}}$) is awarded for successful assignment of HRT flows to AVB. Conversely, if the HRT flow scheduling fails, a negative reward is applied. Additionally, there are cases when assigning HRT flows to TT is not feasible due to constraint violations (as discussed in Section~\ref{sub:constraints}). In such cases, the agent is encouraged to assign HRT flows to AVB to conserve the limited GCL resources. Therefore, if an HRT flow is assigned to TT and fails scheduling, training is halted, signaling to the agent that this is not a viable action.
The agent first assigns HRT flows since they are critical in the network. After the HRT flows are allocated, SRT flow assignment begins, with the objective of maximizing SRT flow utility. To maximize the utility ($U_i(t)$ = 6), the agent tries to assign as many SRT flows as possible to TT. As the HRT flows are already scheduled, only remaining GCL slots are used for SRT flows, when assigned to TT, ensuring that HRT schedulability is unaffected. A positive bonus ($\beta_{\text{bonus}}$) is added to the reward ($r_{\text{SRT, succ}}$) when SRT flows are successfully scheduled with TT. However, if the SRT flow assigned to TT exceeds the buffer deadline of the flow, scheduling is unsuccessful. Our reward design leverages domain knowledge to enable the agent to learn optimal or near-optimal behaviors more efficiently.

\begin{figure}[t!]
    \centering
        \includegraphics[scale=0.19,trim={0.7cm 0.6cm 0.6cm 0.7cm}, clip]{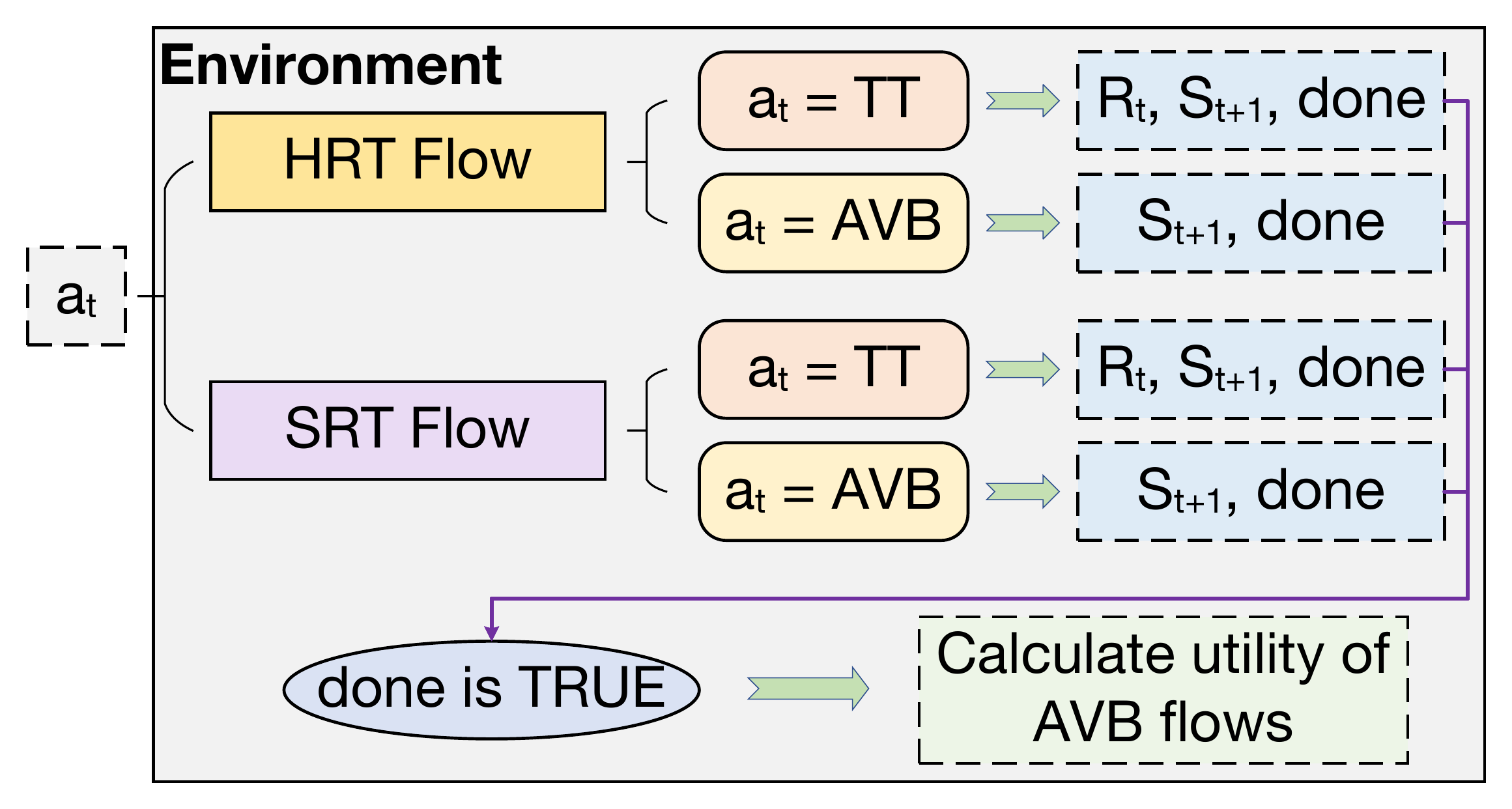}
        \caption{Environment workflow for flow assignment and reward calculation.}
    \label{fig:env}
    \vspace{-0.2cm}
\end{figure}

\begin{algorithm}[t]
\caption{PPO-based TTASeclector}
\begin{algorithmic}[1]
\State \textbf{Input:} $env$, $\gamma$, $\epsilon$, $\eta$, coefficients $c_1$, $c_2$, $\mathbb{R}$, $\text{eval\_freq}$, Link Capacity, flow parameters, topology.
\State \textbf{Output:} Traffic Type Assignment: $\mathcal{TTA}$
\State \textbf{Initialize:} Actor network $\pi_\theta$ with parameters $\theta$, Critic network $V_\phi$ with parameters $\phi$, and Replay Buffer $\mathbb{R}$
\For{each episode $k = 1, 2, \ldots, M$}
    \State Initialize $s_0$, cumulative reward $R_k \gets 0$, $\tau = []$
    \While{episode is not finished}
        \State Sample $a_t \sim \pi_\theta(a|s_t)$ and observe reward $r_t$ and next state $s_{t+1}$
        \State Store $(s_t, a_t, r_t, s_{t+1}, d_t, \pi_\theta(a_t|s_t))$ in trajectory $\tau$
        \State $R_k \gets R_k + r_t$
        \State $s_t \gets s_{t+1}$
    \EndWhile
    \State \textbf{Optimize Actor and Critic} using PPO loss:
    \State Compute action probabilities $\pi_\theta(a_t|s_t)$ and values $V_\phi(s_t)$
    \State Compute surrogate loss: \[
    L^{\text{CLIP}}(\theta) = \mathbb{E}_t \left[ \min \left( \frac{\pi_\theta(a_t|s_t)}{\pi_{\theta_{\text{old}}}(a_t|s_t)} \hat{A}_t, \right. \right. 
    \]
    \[
    \left. \left. \text{clip} \left( \frac{\pi_\theta(a_t|s_t)}{\pi_{\theta_{\text{old}}}(a_t|s_t)}, 1 - \epsilon, 1 + \epsilon \right) \hat{A}_t \right) \right]
    \]
    \State Compute value loss $L^{\text{V}}(\phi) = \frac{1}{2} (G_t - V_\phi(s_t))^2$
    \State Update $\phi$ by minimizing $L^{\text{V}}(\phi)$ using Adam optimizer
    \State Calculate entropy bonus $H(\pi_\theta)$
    \If{$R_k > R_{\text{best}}$}
        \State Save best model
    \EndIf
\EndFor
\end{algorithmic}
\label{alg:ttaselector}
\end{algorithm}

\subsection{Environment}
\label{sub:env}
Fig.~\ref{fig:env} illustrates the behavior of the environment within the model. Once the agent selects an action, the environment then evaluates the criticality of the flow to decide the relevant reward. When the flow is assigned to TT, the environment provides a reward $r_t$ and transitions to the next state ($s_{t+1}$). However, when a flow is assigned to AVB, the model waits until all flow assignments are complete. Once all flows have been assigned a traffic type, the final GCL is computed. The $\mathcal{TTA}$ and $GCL$ are then sent to NetCal to determine the AVB WCD. Subsequently, the reward for AVB flow assignment is calculated based on the WCD and the corresponding utility value. The \textit{done} flag is set to \textit{true} once all flows have been assigned a traffic type. Furthermore, the training is halted if a \textit{STOP} condition appears. 

\subsection{TTASelector Strategy}
\label{sub:ttselect}
Algorithm~\ref{alg:ttaselector} details the PPO algorithm for the TTASelector model. We begin by initializing the actor and critic networks along with the replay buffer (Line 3). Training starts in Line 4, iterating through each episode up to the maximum defined episodes, $M$. In Line 5, we reset the environment to acquire the initial state, initialize the cumulative reward for the episode, and set up the trajectory. Line 7 samples an action from the current policy, yielding a reward and a new state observation. Line 8 stores this observed tuple in the trajectory for later optimization, and Line 9 updates the cumulative reward. In Line 13, the action probabilities and value estimates are computed from the critic network. The PPO objective employs a clipped surrogate loss function, which compares the action probability ratio of the current policy to the old policy, using the minimum of the unclipped and clipped advantage-scaled ratios to stabilize training (Line 14). Lines 15-16 update the critic network by minimizing the difference between estimated values and computed returns. Finally, Line 18-19 compares and saves the model if it achieves a new best performance.

\begin{table}[t!]
\centering
\renewcommand{\arraystretch}{1.1}
\caption{PPO Hyperparameters}
\begin{tabular}{|c|c|}
\hline
\textbf{Parameter} & \textbf{Value} \\
\hline
Learning Rate $\eta$ & 0.0003  \\
Discounted factor $\gamma$ & 0.0001 \\
Entropy $\alpha$ & 0.0003 \\
Clip range $\epsilon$ & 0.2 \\
Value loss function coefficient $c1$ & 0.5 \\
Entropy bonus coefficient $c2$ & 0.02 \\
Evaluation ratio & 4 \\
Episodes per run & 1000 \\
Steps per episode & 900 \\
\hline
\end{tabular}
\label{tab:ppo_hyperparameters}
\end{table}

\subsection{Training}
\label{sub:training}
We trained the model on various network topologies with different numbers of flows. During testing, the model is evaluated on the specific topologies it was trained on but with unseen numbers of flows and flow parameters. We created two distinct sets of test cases, one for training and another for testing. To prevent training bias, we used test cases with entirely different numbers of flows and flow parameters than those seen during training. This approach is particularly advantageous in dynamic environments where the number of flows changes, allowing the model to deliver efficient solutions quickly without requiring retraining.

\section{Evaluation and Discussion} 
\label{sec:results}
For evaluation, we use five synthetic test cases and one realistic test case: the Orion Crew Exploration Vehicle (CEV)~\cite{rubi_ccnc, rubi_noms}. As mentioned earlier, in our model, we work with a single TT queue, single AVB queue, and one AVB class (class A). Additionally, we consider the non-preemptive integration mode~\cite{rubi_noms}. For the synthetic test cases, we generated five different ring topologies with varying numbers of flows, including diverse HRT and SRT flow configurations. We compare the proposed ``TTASelector'' with the baseline ``AVB Only'' flow assignment to emphasize the importance of assigning appropriate traffic types in TSN. It is also possible to assign all the flows to TT, however, we are not comparing the ``TT Only'' solution in our evaluation, as the generation of GCL for large number of flows is computationally expensive. Furthermore, as discussed in \cite{voica_traffic_assignment}, with large number of TT flows in the network, the $\mathcal{ES}$ and $\mathcal{SW}$ may run out of memory for the GCL. Our results shows that ``TTASelector'' can schedule a larger number of HRT flows while also improving the utility and QoS of SRT flows. The parameters used in our model are listed in Table.~\ref{tab:ppo_hyperparameters}. All experiments and simulations were conducted on a Windows laptop equipped with an Intel\textregistered\ Processor Core\texttrademark\ i7-10610U CPU running at 1.80GHz, and 32 GB RAM. 

\begin{figure}[t!]
    \centering
    \begin{minipage}[t]{0.24\textwidth}
        \centering
        \includegraphics[scale=0.29,trim={0.35cm 0.2cm 0.3cm 0.2cm}, clip]{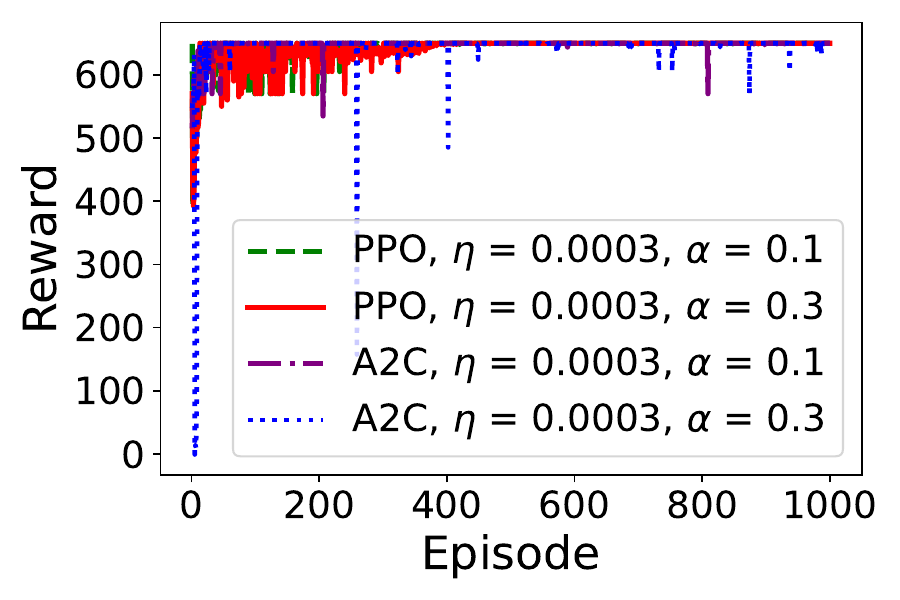}
        \subcaption[]{}
        \label{fig:ppo_a2c_train}
    \end{minipage}%
    \hfill
    \begin{minipage}[t]{0.24\textwidth}
        \centering
        \includegraphics[scale=0.29,trim={0.35cm 0.2cm 0.3cm 0.2cm}, clip]{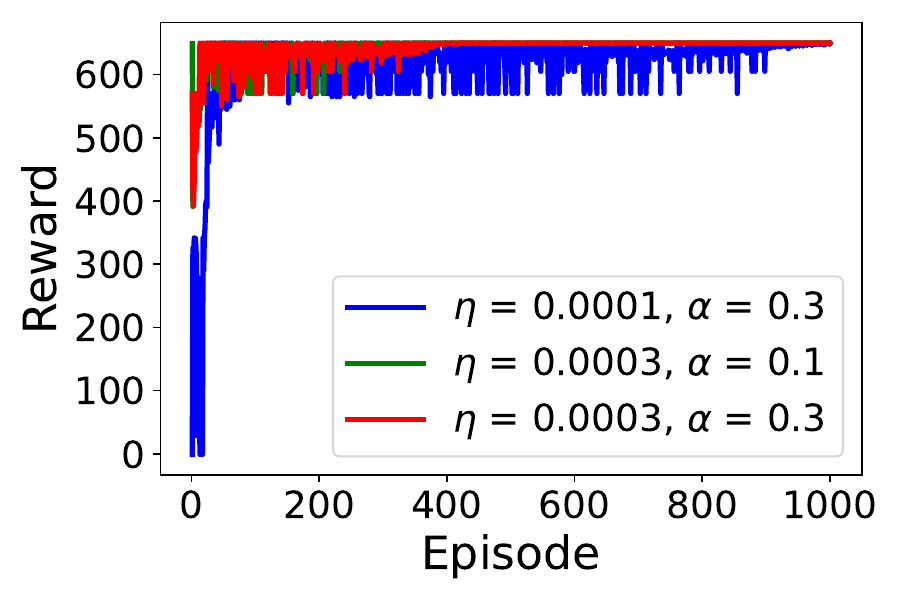}
        \subcaption[]{}
        \label{fig:ppo_reward_train}
    \end{minipage}
    \caption{Convergence curves: (a) PPO and A2C based models showing the reward during training, and (b) performance of different parameters of PPO during training.}
    \label{fig:train}
\end{figure}

Fig.~\ref{fig:ppo_a2c_train} presents a comparison of training rewards for the Advantage Actor-Critic (A2C) and PPO-based TTASelector model across 1000 episodes. PPO exhibited better performance in our model, with greater training stability, leading to our decision to use PPO in the model. For hyperparameter tuning, we compared the PPO model with different parameter configurations, as shown in Fig.~\ref{fig:ppo_reward_train}. We compare the schedulability and computational time cost for the realistic test case using TTASelector against the method presented in \cite{voica_traffic_assignment}. Table~\ref{tab:orion_time_cost} shows the TTA of HRT and SRT flows for the Orion topology. The metaheuristic algorithm successfully assigned 94.94\% of HRT flows, whereas TTASelector achieved 100\% schedulability for HRT flows, though with some compromise in the schedulability of SRT flows. Most notably, the runtime for the metaheuristic algorithm exceeded 12 hours, while TTASelector completed its TTA in approximately 3 seconds (testing time). It is important to note that the training time of TTASelector is different from the testing time. We do not consider the training time in our evaluation, as the model is trained once, stored, and later used directly for testing. 

Table ~\ref{tab:tc_result} presents results for various test cases across different topology sizes and flow numbers. When all HRT and SRT flows are assigned as AVB, both HRT and SRT schedulability rates decrease, occasionally dropping to 50\%. The TTASelector model optimizes traffic type assignments, performing assignments that maximize overall network schedulability. With TTASelector, schedulability reaches up to on an average 95\%, and runtime remains under 4 seconds.

\begin{table*}[h]
\caption{Comparison of ``AVB only'' vs. ``TTASelector'' showing the improvement in the schedulability for both HRT and SRT flows. The runtime of ``TTASelector'' further demonstrates the computation time cost of our model.}
\centering
\renewcommand{\arraystretch}{1.2}
\begin{tabular}{|c|cc|cc|cc|cc|c|}
\hline
\multirow{2}{*}{ID}  & \multicolumn{2}{c|}{Number of Flows} & \multicolumn{2}{c|}{Topology Size} & \multicolumn{2}{c|}{AVB} & \multicolumn{2}{c|}{TTASelector} & \multicolumn{1}{c|}{Avg. Runtime}                 
\\ \cline{2-9} & \multicolumn{1}{c|}{HRT} & SRT & 

\multicolumn{1}{c|}{\begin{tabular}[c]{@{}c@{}}ES\end{tabular}} & \begin{tabular}[c]{@{}c@{}}SW\end{tabular} & \multicolumn{1}{c|}{\begin{tabular}[c]{@{}c@{}}HRT scheduled\end{tabular}} & \begin{tabular}[c]{@{}c@{}}SRT utility\end{tabular} & \multicolumn{1}{c|}{\begin{tabular}[c]{@{}c@{}}HRT scheduled\end{tabular}} & \begin{tabular}[c]{@{}c@{}}SRT utility\end{tabular}  & \multicolumn{1}{c|}{\begin{tabular}[c]{@{}c@{}}[in secs]\end{tabular}} \\ \hline

\multirow{3}{*}{TC1} & \multicolumn{1}{c|}{59}    & 174   & \multicolumn{1}{c|}{25} & 6 & \multicolumn{1}{c|}{88.13\%} & 91.09\% & \multicolumn{1}{c|}{100\%} & 94.00\% & 2.3  \\ 

\cline{2-10} & \multicolumn{1}{c|}{78} & 155 & \multicolumn{1}{c|}{25} & 6 & \multicolumn{1}{c|}{82.05\%} & 93.00\% & \multicolumn{1}{c|}{100\%} & 96.13\% & 2.3 \\ 

\cline{2-10} & \multicolumn{1}{c|}{117} & 116 & \multicolumn{1}{c|}{25} & 6 & \multicolumn{1}{c|}{57.26\%} & 64.67\% & \multicolumn{1}{c|}{100\%} & 92.52\% & 2.6 \\ \hline

\multirow{3}{*}{TC2} & \multicolumn{1}{c|}{87} & 172 & \multicolumn{1}{c|}{35} & 9  & \multicolumn{1}{c|}{96.55\%} & 94.56\% & \multicolumn{1}{c|}{100\%} & 97.12\%  & 2.7 \\ 

\cline{2-10} & \multicolumn{1}{c|}{104} & 208 & \multicolumn{1}{c|}{35} & 9 & \multicolumn{1}{c|}{92.30\%} & 92.69\% & \multicolumn{1}{c|}{100\%} & 95.15\%  & 2.98 \\ 

\cline{2-10} & \multicolumn{1}{c|}{282} & 281   & \multicolumn{1}{c|}{35} & 9 & \multicolumn{1}{c|}{43.62\%} & 51.21\% & \multicolumn{1}{c|}{99.29\%} & 92.62\% & 3.01 \\ 

\hline

\multirow{3}{*}{TC3} & \multicolumn{1}{c|}{76} & 151  & \multicolumn{1}{c|}{52} & 13  & \multicolumn{1}{c|}{92.10\%} & 96.61\% & \multicolumn{1}{c|}{100\%} & 98.30\%  & 2.87 \\ 

\cline{2-10} & \multicolumn{1}{c|}{100} & 200 & \multicolumn{1}{c|}{52} & 13 & \multicolumn{1}{c|}{93.00\%} & 94.10\% & \multicolumn{1}{c|}{100\%} & 96.00\%  & 2.99 \\ 

\cline{2-10} & \multicolumn{1}{c|}{187} & 187   & \multicolumn{1}{c|}{52} & 13 & \multicolumn{1}{c|}{65.24\%} & 68.23\% & \multicolumn{1}{c|}{99.36\%} & 90.50\% & 3.01\\ 

\hline
\multirow{3}{*}{TC4} & \multicolumn{1}{c|}{100}   & 200 & \multicolumn{1}{c|}{77} & 18 & \multicolumn{1}{c|}{93.00\%} & 94.10\% & \multicolumn{1}{c|}{100\%} & 96.00\% & 3.05\\ 

\cline{2-10} & \multicolumn{1}{c|}{150} & 150 & \multicolumn{1}{c|}{77} & 18 & \multicolumn{1}{c|}{92.30\%} & 93.16\% & \multicolumn{1}{c|}{100\%} & 96.28\%  & 3.1 \\ 

\cline{2-10} & \multicolumn{1}{c|}{226} & 226   & \multicolumn{1}{c|}{77} & 18 & \multicolumn{1}{c|}{61.06\%} & 61.44\% & \multicolumn{1}{c|}{98.67\%} & 95.35\% & 3.13 \\ 

\hline
\multirow{3}{*}{TC5} & \multicolumn{1}{c|}{86}    & 171 & \multicolumn{1}{c|}{104} & 24 & \multicolumn{1}{c|}{81.39\%} & 92.62\% & \multicolumn{1}{c|}{100\%} & 95.68\%  & 3.01\\ 

\cline{2-10} & \multicolumn{1}{c|}{135} & 269   & \multicolumn{1}{c|}{104} & 24 & \multicolumn{1}{c|}{78.51\%} & 88.03\% & \multicolumn{1}{c|}{100\%} & 96.20\%  & 3.39\\ 

\cline{2-10} & \multicolumn{1}{c|}{269} & 135   & \multicolumn{1}{c|}{104} & 24 & \multicolumn{1}{c|}{63.19\%} & 59.75\% & \multicolumn{1}{c|}{95.91\%} & 91.12\% & 3.41 \\ 

\hline
\end{tabular}
\label{tab:tc_result}
\end{table*}

\begin{table*}[h]
\caption{Comparison of ``Tabu Search-based metaheuristic''~\cite{voica_traffic_assignment} with our ``TTASelector'' model for realistic test case ``Orion''.}
\centering
\renewcommand{\arraystretch}{1.2}
\begin{tabular}{|l|l|l|l|l|l|l|l|l|l|l|}
\hline
& \multicolumn{2}{c|}{Number of Flows} & \multicolumn{2}{c|}{Topology Size} & \multicolumn{3}{c|}{Metaheuristic} & \multicolumn{3}{c|}{TTASelector} \\ 
\hline
\multirow{2}{*}{Orion} & \multicolumn{1}{l|}{HRT}   & SRT & \multicolumn{1}{l|}{ES} & SW  & \multicolumn{1}{l|}{\begin{tabular}[c]{@{}l@{}}HRT scheduled \end{tabular}} & \begin{tabular}[c]{@{}l@{}}SRT utility \end{tabular} & \begin{tabular}[c]{@{}l@{}}Runtime[h:mm]\end{tabular} & \multicolumn{1}{l|}{\begin{tabular}[c]{@{}l@{}}HRT scheduled\end{tabular}} & \begin{tabular}[c]{@{}l@{}}SRT utility\end{tabular} & \begin{tabular}[c]{@{}l@{}}Runtime [secs]\end{tabular} \\ 

\cline{2-11} 
& \multicolumn{1}{l|}{99}  & 87 & \multicolumn{1}{l|}{31} & 15 & \multicolumn{1}{l|}{{\cellcolor{red!40}94.94\%} $\boldsymbol{\downarrow}$} & 98.68\% & {\cellcolor{red!40}12:30 $\boldsymbol{\uparrow}$} & \multicolumn{1}{l|}{\cellcolor{green!30}100\% $\boldsymbol{\uparrow}$}  & 95.98\% & {\cellcolor{green!30}3.01 $\boldsymbol{\downarrow}$} \\ \hline
\end{tabular}
\label{tab:orion_time_cost}
\vspace{-0.3cm}
\end{table*}

\section{Conclusion}
\label{sec:conclusion}
In this paper, we addressed the traffic type assignment problem in a mixed-critical TSN network. Our results demonstrate that a learning-based approach provides an efficient and faster solution for traffic type assignment compared to traditional methods. We presented an end-to-end model that integrates configuration, network calculus analysis, and scheduling generation within a single framework. The TTASelector model, based on a reinforcement learning-based algorithm, enables traffic type assignment in TSN networks, reducing the need for manual intervention by system engineers. Such models lay the groundwork for future advancements, including the integration of learning based models for TSN configuration.  

\section{Acknowledgment}
\label{sec:ack}
We thank Zhen Yu for his contribution with the initial modeling work, which influenced parts of this paper.

\renewcommand{\baselinestretch}{0.95}
\bibliographystyle{IEEEtran}

\bibliography{reference}
\end{document}